\documentclass[11pt]{article}
\usepackage{bm,graphicx}

\begin{document}

\title{Phase-Induced (In)-Stability in Coupled Parametric
Oscillators}

\author{Mauro Copelli and Katja Lindenberg\\
Department of Chemistry and Biochemistry 0340\\
University of California San Diego\\
La Jolla, California 92093-0340}

\date{\today}
\maketitle

\begin{abstract}
We report results on a model of two coupled oscillators that undergo
periodic parametric modulations with a phase difference
$\theta$. Being to a large extent analytically solvable, the model
reveals a rich $\theta$ dependence of the regions of parametric
resonance. In particular, the intuitive notion that anti-phase
modulations are less prone to parametric resonance is confirmed for
sufficiently large coupling and damping. We also compare our results
to a recently reported mean field model of collective parametric
instability, showing that the two-oscillator model can capture much of
the qualitative behavior of the infinite system.

\vspace{8pt}
\vspace{8pt}

\end{abstract}

\section{Introduction}
\label{sec:intro}

Parametric resonance is a phenomenon pervading several fields of
science. It occurs when the modulation of a system parameter causes
the system to become unstable. The literature on parametric resonance is
enormous, so the best one can do is cite a range of different
subjects where it may play a role. Examples include mechanical systems
where such resonances were first identified~\cite{Landau,Arnold,Nayfeh,Bena},
elementary particles~\cite{neutrinos}, quantum dots~\cite{dots},
astrophysics~\cite{astrophysics}, fluid mechanics~\cite{fluid}, plasma
physics~\cite{plasma}, electronic networks~\cite{electronic},
superconducting and laser devices~\cite{laser},
biomechanics~\cite{biomechanics}, and even medicine~\cite{medicine}.
Connections with chaotic systems have been suggested
recently~\cite{Kobes00}.
The simplest and perhaps most familiar example of parametric resonance
occurs in a harmonic oscillator whose frequency varies
periodically with time~\cite{Landau,Arnold,Nayfeh,Bena}. For certain
ranges of modulation parameters (frequency, amplitude) the oscillator
is unstable while for others it is stable.  Even for this seemingly simple
system the stability boundary diagram is already quite rich and
complex (see below).

A great deal of recent work has dealt with systems of {\em coupled}
oscillators -- again, the literature in this general area is enormous.
However, very little attention has focused on systems of coupled parametric
oscillators~\cite{Bena99b,Ji98}. Such coupled arrays are
particularly intriguing
because each single oscillator alone exhibits regions of stable or
unstable behavior. Two questions arise naturally: 1) How does 
coupling modify the single-oscillator stability boundaries? 2) Are there
collective parametric instabilities of the coupled system that are
distinct from those experienced by single oscillators?
In a recent paper Bena and Van den Broeck~\cite{Bena99b} addressed these
questions for an infinite set of globally mean-field coupled harmonic
oscillators subjected to time-periodic block pulses with quenched uniformly
distributed random phases (``quenched" in this context means that the phase
of frequency modulation of each oscillator is set at time $t=0$ and
then remains unchanged).  Within their mean field treatment they are
able to deal with both of the questions posed above. In particular,
they find wide ranges of parameter values that lead to collective
instabilities.

Our interest in this paper is to identify generic coupling effects that are
not specifically a consequence of the mean field analysis, and in this context
to understand how a coupled system of (very) few oscillators with
short-ranged interactions might carry in it the seeds of the
infinite/infinitely cross-coupled array.  In particular, we seek the
seeds of the collective parametric instabilities.
We do this by studying a model of
two coupled oscillators.  The control parameter in our study is the phase
difference in the periodic modulation of the frequencies of the two
oscillators.

In Section~\ref{sec:model} we introduce the model and its
partially analytic solution.
Section~\ref{sec:results} is a presentation of results
as a large collection of stability boundary diagrams that convey the
way in which stability boundaries shift, appear, and disappear
with parameter changes.  In
Section~\ref{sec:collective} we analyze the ways in which the seeds of
the collective parametric instabilities found in the mean field model
already appear in the two-oscillator system.  A brief summary of our
conclusions is reiterated in Section~\ref{sec:conclusions}.

\section{The model}
\label{sec:model}

Our system consists of two linearly coupled parametric oscillators whose
equations of motion are
\begin{eqnarray}
\label{eq:twoosc}
\ddot{x}_{1} & = & -\omega_{0}^2[1+\phi_{1}(t)]x_{1} - k(x_{1}-x_{2})
-\gamma \dot{x}_1
\nonumber \\ [12pt]
\ddot{x}_{2} & = & -\omega_{0}^2[1+\phi_{2}(t)]x_{2} - k(x_{2}-x_{1})
-\gamma \dot{x}_2
\; .
\end{eqnarray}
Here $\omega_{0}$ is the natural frequency of each (uncoupled)
oscillator, $k$ is the coupling constant between them and $\gamma$ is the
damping coefficient. The parametric modulations $\phi_{1}(t)$ and
$\phi_{2}(t)$ are square waves with period $T\equiv 2\pi/\omega_{p}$,
amplitude $A$ and phase difference $\theta$:
\begin{eqnarray}
\label{eq:phasedif}
\phi_{1}(t) & = & A\;\mbox{sgn}\left(\sin(\omega_{p}t)\right)
\nonumber \\[12pt]
\phi_{2}(t) & = & A\;\mbox{sgn}\left(\sin(\omega_{p}t+\theta)\right)
\; .
\end{eqnarray}
With $A=0$ one is left with coupled ordinary damped harmonic
oscillators whose total energy decays exponentially to zero.
In the presence of the parametric
modulation, however, energy is periodically pumped into the system,
which may or may not lead to parametric resonance, i.e., to an infinite
growth of the amplitudes of the oscillators. Our goal is to determine the
boundaries between these two behaviors, which will be referred to
as unstable and stable.  As a point of reference we point the reader to
the appendix where we review the results for a single 
parametric oscillator.

It is worth noting that
a simple rescaling of time to dimensionless units, $t^\prime = t\omega_0$,
shows that Eqs.~(\ref{eq:twoosc}) are governed by the dimensionless
parameter combinations $r \equiv \omega_{0}/\omega_{p}$, $A$,
$k/\omega_{0}^2$, $\gamma/\omega_{0}$ and $\theta$. Moreover, the
behavior of the system
is invariant with respect to the replacement of $\theta$ by
$\theta^\prime \equiv 2\pi - \theta$ (reflection around $\pi$) since
this just amounts to an exchange of indices between the oscillators.

\subsection{Floquet theory}
\label{sec:floquet}

The linearity of the equations allows one to make use of Floquet
theory to solve the problem~\cite{Nayfeh}.  Defining $\bm{X} \equiv
(x_{1},\dot{x}_{1},x_{2},\dot{x}_{2})^{\rm T}$ where the superscript
${\rm T}$ denotes
the transpose (i.e., $\bm{X}$ is a column matrix), one can rewrite
Eqs.~(\ref{eq:twoosc}) in matrix form as $\dot{\bm{X}}(t) = \hat{D}(t)
\bm{X}(t)$ with a matrix $\hat{D}$ satisfying $\hat{D}(t) =
\hat{D}(t+T)$. Given a solution $\bm{X}(t)$ to
Eqs.~(\ref{eq:twoosc}), the time periodicity of the parametric
modulations thus implies that $\bm{X}(t+T)$ is also a
solution. Therefore one can find a matrix $\hat{F}$ such that
$\bm{X}(t+T) = \hat{F}(T)\bm{X}(t)$ and hence more generally through a
repetition of this solution $\bm{X}(t+nT) = \hat{F}^n(T)\bm{X}(t)$.
The long time behavior of
the system is thus clearly determined by the eigenvalues
$\{\lambda_{i}\}$ of the {\em Floquet operator\/} $\hat{F}$, which
propagates the system in phase space for one period of the
modulation. The eigenvectors $\bm{v}_{j}$ of $\hat{F}$ satisfy
$|\bm{v}_{j}(t+nT)| = |\lambda_{j}|^{n}|\bm{v}_{j}(t)|$, so in the
limit $n\to\infty$ parametric resonance occurs if
$\max_{j}\{|\lambda_{j}|\}>1$.

The model is appealing because for any number of
oscillators (a single oscillator, or the coupled oscillators considered
here, or even the mean field version of an infinitely cross-coupled
infinite chain~\cite{Bena99b}) the piecewise constant parametric
modulation leads to a piecewise linear system whose piecewise
solution is known analytically.
One can therefore construct the Floquet operator explicitly by
simply multiplying together {\em
piecewise linear Floquet operators\/}. This provides an immense
reduction in computational effort by skipping what is usually the most
time-consuming task in obtaining the regions of parametric resonance,
namely, the numerical evaluation of the Floquet operator itself.
For an isolated oscillator this procedure is well
known~\cite{Arnold,Bena} and reviewed in the appendix.

For the sake of clarity, let us first analyze the
frictionless case $\gamma=0$. Each linear interval is
characterized by one of the four possible states of the modulations
$(\phi_1,\phi_2)= (+A,+A)$, $(+A,-A)$, $(-A,+A)$, or $(-A,-A)$.
For each of these
states, the solutions are of the form $X_{j}(t) = A_{j}e^{iPt} +
B_{j}e^{iMt}$, where the eigenfrequencies $P$ and $M$ follow directly
from the diagonalization of $\hat{D}$:
\begin{eqnarray}
P^{2} & = & \frac{\omega_{1}^2+\omega_{2}^2 +
\sqrt{(\omega_{1}^2-\omega_{2}^2)^2 + 4k^2}}{2} \nonumber \\ [12pt]
M^{2} & = & \frac{\omega_{1}^2+\omega_{2}^2 -
\sqrt{(\omega_{1}^2-\omega_{2}^2)^2 + 4k^2}}{2}\; ,
\end{eqnarray}
where $\omega_{1,2}^2 \equiv
\omega_{0}^2[1+\phi_{1,2}(t)]+k$. Denoting the current state of the
modulation by indices $+$ and $-$, notice that while $P_{++}$,
$M_{++}$ and $P_{+-}=P_{-+}$ are always real, $M_{--}$ becomes
imaginary when $A>1$ while $P_{--}$ and $M_{+-}=M_{-+}$ become
imaginary when $A > 1+2k/\omega_{0}^2$. Whether the system
periodically alternates between harmonic behavior and one or more
saddle nodes will thus depend on the parameter region and the phase
difference $\theta$.

During a time interval $\tau$ with fixed $\phi_1,\phi_2$ we can
relate $\bm{X}(t+\tau)$ to $\bm{X}(t)$ as
$\bm{X}(t+\tau)= \hat{f}(\tau)\bm{X}(t)$, where $\hat{f}$ is
the {\em piecewise} linear Floquet operator:
\begin{equation}
\label{eq:pieceFloquet}
\hat{f}(t) =
%\bm{X}(t)  =
%\underbrace
\frac{1}{P^2-M^2}
\left(
\begin{array}{cccc}
-m_1 c_p + p_1 c_m & -m_1 s_p + p_1 s_m &
k[- c_p + c_m] & k[- s_p + s_m] \\ 
m_1 P^2 s_p - p_1 M^2 s_m & -m_1 c_p + p_1 c_m &
k[P^2 s_p - M^2 s_m] & k[-c_p + c_m] \\ 
k[-c_p + c_m] & k[-s_p + s_m] &
-m_2 c_p + p_2 c_m & -m_2 s_p + p_2 s_m \\ 
k[P^2 s_p - M^2 s_m] & k[-c_p + c_m] &
m_2 P^2 s_p - p_2 M^2 s_m & -m_2 c_p + p_2 c_m
\end{array}
\right)
%}_{\displaystyle\equiv \hat{f}(t)}
%\bm{X}(0) \; ,
\end{equation}
where we make use of the shorthand notation $c_p \equiv \cos(Pt)$,
$s_p \equiv P^{-1}\sin(Pt)$, $c_m \equiv \cos(Mt)$, $s_m \equiv
M^{-1}\sin(Mt)$, $m_{1,2} \equiv M^2 - \omega_{1,2}^2$ and $p_{1,2}
\equiv P^2 - \omega_{1,2}^2$. The Floquet operator is then finally
obtained as the product of piecewise Floquet operators with
arguments that depend on the
phase difference between the modulations:
\begin{equation}
\hat{F}(T) =
\hat{f}_{-+}\left(\frac{T}{2}\frac{\theta}{\pi}\right)
\hat{f}_{--}\left(\frac{T}{2}\left[1-\frac{\theta}{\pi}\right]\right)
\hat{f}_{+-}\left(\frac{T}{2}\frac{\theta}{\pi}\right)
\hat{f}_{++}\left(\frac{T}{2}\left[1-\frac{\theta}{\pi}\right]\right)
\; .
\end{equation}
The stability properties of the coupled system are determined by
the magnitudes of the four eigenvalues $\lambda_1,\dots,\lambda_4$
of the Floquet operator.
Note that if $\theta = 0$ or $\theta =\pi$, the above expression
reduces to the product of only two matrices (as it should) since
$\hat{f}(0)$ is just the identity matrix. The addition of damping can
be dealt with by noticing that if $\bm{X}(t)$ is a solution for
$\gamma = 0$, then $\bm{Y}(t) \equiv e^{-\gamma t/2}\tilde{\bm{X}}(t)$
is a solution for $\gamma \neq 0$, where $\tilde{\bm{X}}(t) =
\left.\bm{X}(t)\right|_{\omega_{1,2}^2 \to \omega_{1,2}^2 -
\gamma^2/4}$.

\section{Results}
\label{sec:results}

In the absence of damping, Liouville's theorem insures that $\det
\hat{f} = 1$, which is confirmed by cumbersome calculations and in
turn implies $\lambda_1\lambda_2\lambda_3\lambda_4 = \det\hat{F} = 1$.
In contrast with the single oscillator (see appendix), here this
is not a
sufficient condition to guarantee that the parametric resonance
bifurcations occur at $\lambda_{j} = \pm
1$~\cite{Arnold,Bena}: for coupled oscillators
the largest eigenvalue can cross the unit circle $|\lambda_{j}|=1$
and hence enter a region of instability (parametric resonance) in
other directions in the complex plane. The fourth order
characteristic polynomial needs to be solved numerically to
determine the instability boundaries; this is a computationally
inexpensive procedure since we have an analytic expression for the
Floquet operator.

\subsection{In-phase modulations}
\label{sec:thetazero}

The case $\theta=0$ corresponding to in-phase modulation
$\phi_{2}(t)=\phi_{1}(t)$ simplifies sufficiently to
yield analytic results for the stability boundaries. This is the simplest
situation to be studied, since it is possible to reduce the original
four-dimensional problem in this case to two known two-dimensional
systems. This
decoupling is accomplished by changing the coordinate system to the
reference frame of the center of mass. Defining $x \equiv
(x_{1}+x_{2})/2$ and $\rho \equiv (x_{1}-x_{2})/2$, one obtains from
Eq.~(\ref{eq:twoosc})
\begin{eqnarray}
\label{eq:xbar}
\ddot{x} & = & -\omega_{0}^2[1+\phi_{1}(t)]x -
\gamma\dot{x} \\ [12pt]
\label{eq:rho}
\ddot{\rho} & = &  -\left\{\omega_{0}^2[1+\phi_{1}(t)]+2k\right\}\rho
- \gamma\dot{\rho}
\; .
\end{eqnarray}
Each of these equations is precisely that of a single parametric
oscillator, one with frequency $\omega_0$ and modulation amplitude
$A$, the other with rescaled parameters
\begin{figure}[!htb!]
\begin{center}
\includegraphics[width = 0.7\textwidth, angle = -90]{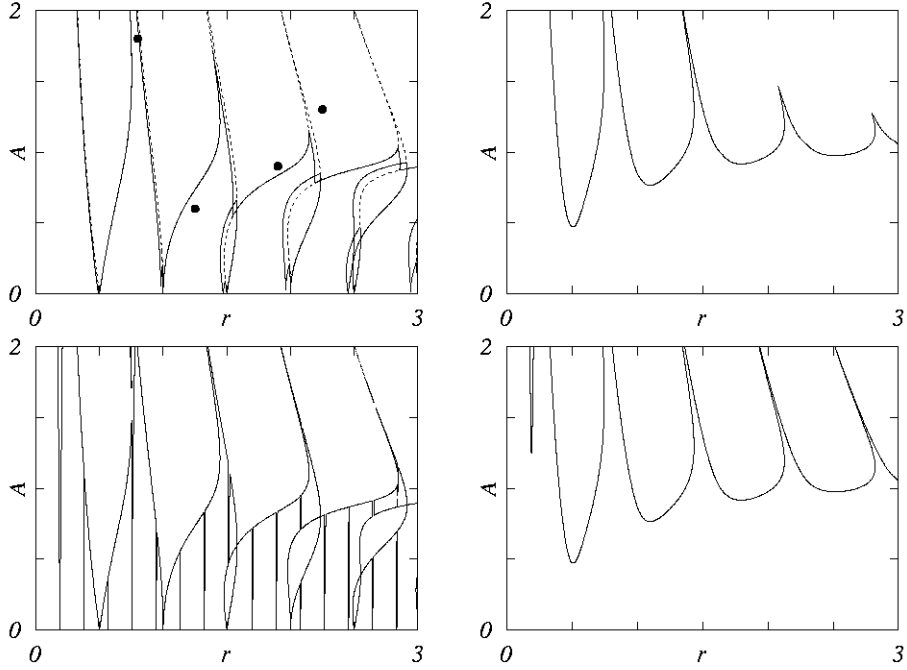}
\caption{Instability boundaries for $\theta = 0$ in the $(r,A)$ plane:
$\gamma = 0$ in the first column while $\gamma/\omega_{0} = 0.3$ in
the second column.  $k/\omega_{0}^2 = 0.02$ in the first row, while
$k/\omega_{0}^2 = 3$ in the second row.  The dashed line in the first
panel is the result for $k = 0$ (single uncoupled oscillator). The
dark points in this panel are parameter pairs to be considered in more
detail subsequently.}
\label{fig:theta0}
\end{center}
\end{figure}
\begin{eqnarray}
\label{eq:omegaprime}
\omega^{\prime}_{0} & \equiv & \sqrt{\omega_{0}^2 + 2k}
\nonumber \\ [12pt]
A^\prime & \equiv & A(\omega_{0}/\omega^{\prime}_{0})^2\; .
\end{eqnarray}
We can thus use the known results for the single
oscillator~\cite{Arnold,Bena}, for which a closed expression
exists for the boundary of the instability regions in the $(r,A)$
plane (see the appendix).

For an undamped parametric oscillator with parameters $(r,A)$
parametric instabilities occur inside ``tongues" that open up from
integer and half-integer values of $r$ as discussed in the
appendix and shown by the dashed curves in the first panel of
Fig.~\ref{fig:theta0}.  For the coupled problem in which we have
transformed the problem to two independent parametric oscillators
with different parameters, the instability regions are given by
the overlap of the sets of tongues arising from each independent oscillator,
one set emerging from integer and half-integer values of $r$
(``$r$-instability regions"), and the other from
\begin{equation}
\label{eq:rprime}
r^{\prime}(k) \equiv r\sqrt{1+2k/\omega_{0}^2} =
\frac{\omega_{0}^{\prime}}{\omega_{p}}
\end{equation}
(``$r'$-instability regions").
When the coupling $k$ between the oscillators is small then the
two sets of tongues almost overlap and one obtains in the undamped
case the solid curves in the first panel, which show the
instability boundaries when $k/\omega_0^2=0.02$.  Still in the
undamped problem, when the coupling $k$ is large the two sets of
tongues occur at different parameter scales. The boundary
diagram for $k/\omega_0^2=3$ is shown in lower left panel of the
figure.  In either case, since now we have two independent
sets of instability regions the effect of the coupling $k$ for
$\theta=0$ has been to {\em enlarge} the parametric resonance regime
relative to that of two uncoupled oscillators.

This last result is completely general: the effect of coupling for
in-phase parametric modulations is to {\em enlarge} the parametric
resonance regime relative to the uncoupled case {\em regardless} of
the form of the periodic modulation.  The details of the phase
boundaries will of course depend on the specific modulation function.

Damping, even in an isolated parametric oscillator, destroys much of
the intricate boundary structure and increases the regions of
stability.  This is also the case for coupled oscillators, where
increasing the damping tends to blur out the effects of coupling.
These behaviors can be seen in the right column of
Fig.~\ref{fig:theta0}.  The regions of instability are now restricted
to larger amplitudes $A$. Note the similarity between these two
figures, which involve different couplings but now with substantial
damping, $\gamma/\omega_{0} = 0.3$. Note also that in the lower panel
of this column (large coupling) the very first instability wedge is an
$r'$-instability region while the other portions of the diagram (as
well as the unstable regions in the upper weak-coupling case) include
both $r$ and $r'$ instabilities.

For the particular case $\theta=0$ these
results allow us to say something about the interesting problem of
asymptotic {\em synchronization}.
In regions where the center of mass coordinate
$x$ is unstable but the relative coordinate $\rho$ is stable
($r$-instability regions that do not overlap with $r'$-instability
regions) the
coupled oscillators are synchronized if $\gamma\neq 0$
($x_1=x_2$), that is, the
two oscillators move together about the origin with ever increasing
amplitude.  Conversely, if
$x$ is stable but $\rho$ is unstable ($r'$-instability regions that do not
overlap with $r$-instability regions), with $\gamma\neq 0$ 
the oscillators become ``antisynchronized" ($x_1=-x_2$), that is, the two
oscillators oscillate with ever increasing amplitude but in opposite
directions, crossing one another each time they pass through the origin.
Antisynchronization becomes more difficult to achieve with
increasing coupling.
If $\gamma=0$ the strict equalities $x_1=x_2$ or $x_1=-x_2$
no longer hold
in the non-overlapping regions, but the difference between $x_1$
and $x_2$ or $-x_2$ is oscillatory and remains bounded.
A simultaneous instability of both $x$ and $\rho$ involves an unstable
center of mass coordinate and unbounded oscillations of each oscillator
about this unbounded mean, which in turn involves a more
complicated phase relation between the motions of the two oscillators.
We return to this issue later.

\subsection{Out-of-phase modulations}
\label{sec:thetagen}

A wealth of very intricate results arises when $\theta\neq 0$. In
contrast with the $\theta=0$ case, it is now no longer clear how to
break down the problem into simpler independent components (even for
$\theta=\pi$) and, in particular, there is no longer a transparent way
to relate the results to those of single parametric oscillators.

\begin{figure}[!htb!]
\begin{center}
\includegraphics[width = 0.8\textwidth, angle = 0]{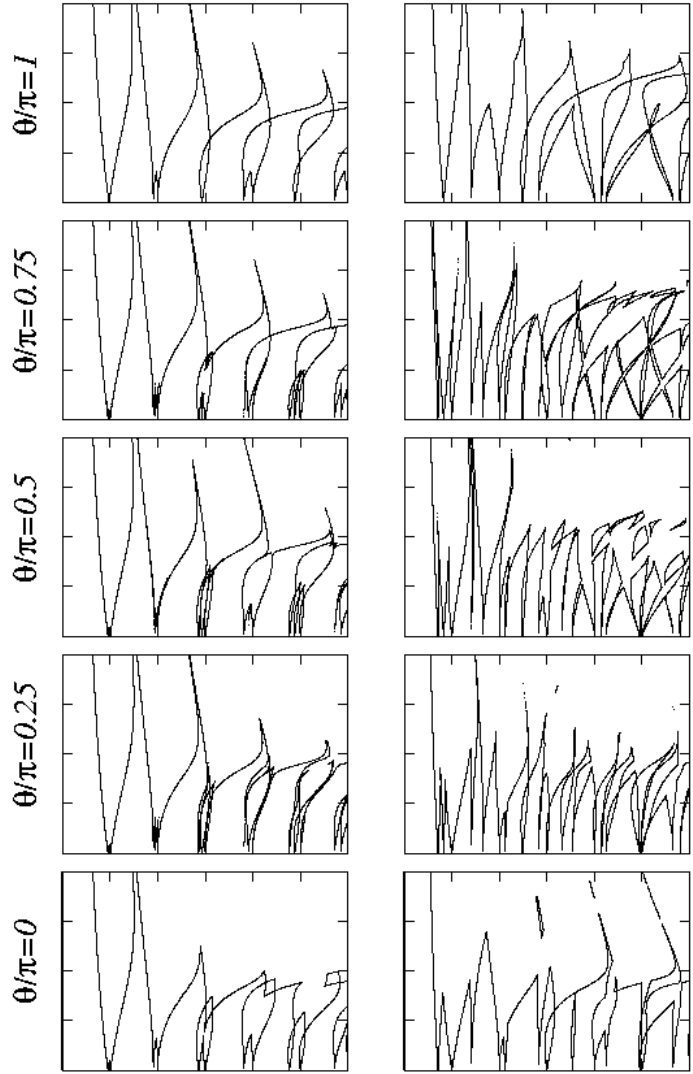}
\caption{$(r,A)$ plane as in Fig.~\ref{fig:theta0} (origin is at
bottom left corner; tic marks are separated by half a unit on each
axis). Left column: $k/\omega_{0}^2 = 0.05$. Right column:
$k/\omega_{0}^2 = 0.5$. $\gamma = 0$ for all panels.}
\label{fig:gamma0}
\end{center}
\end{figure}

\begin{figure}[htb!]
\begin{center}
\includegraphics[width = 0.7\textwidth, angle= -90]{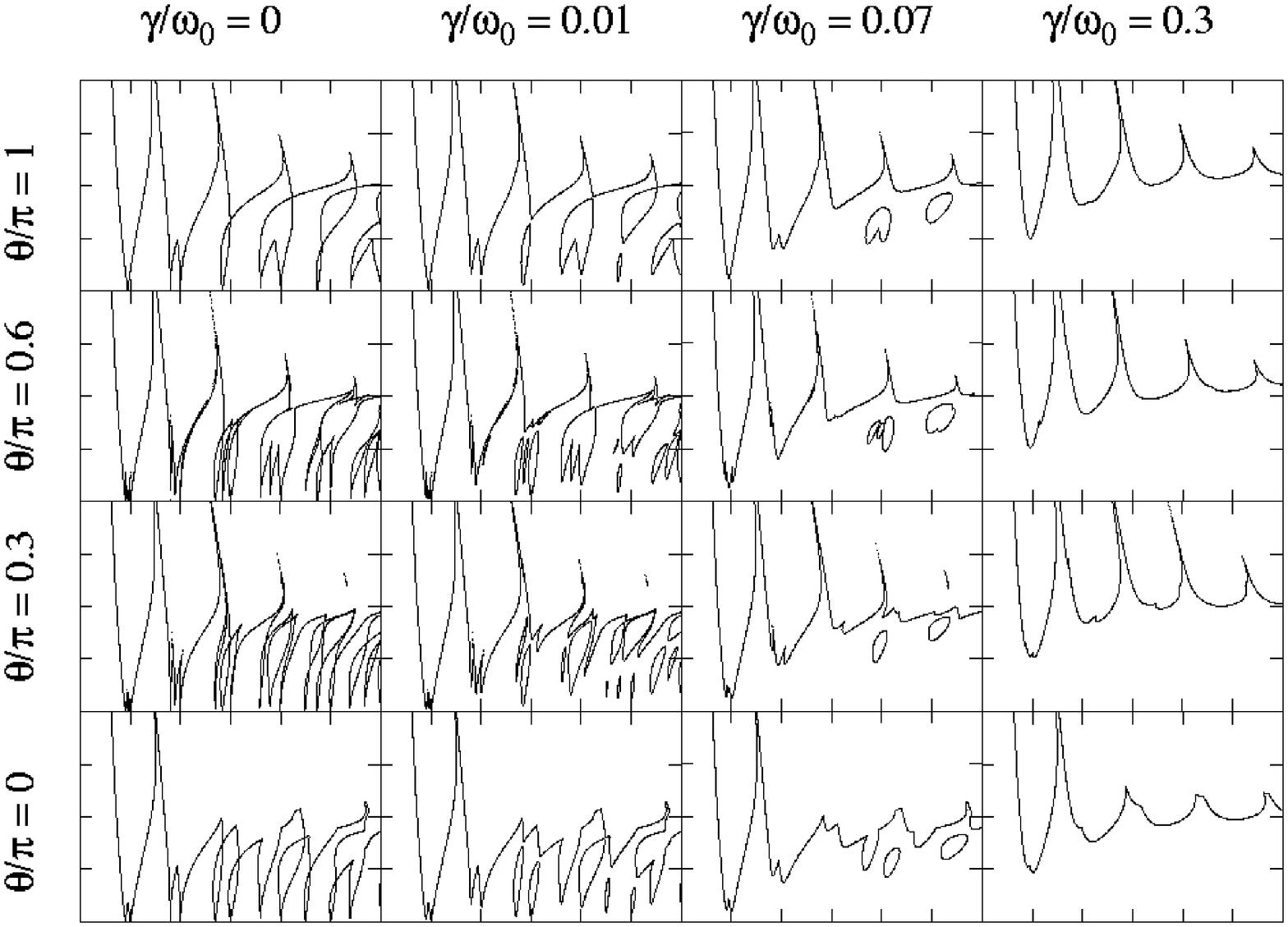}
\caption{$(r,A)$ plane as in Fig.~\ref{fig:theta0} for $k/\omega_{0}^2
= 0.12$ and several combinations of $\theta$ and $\gamma$ (origin is
at bottom left corner; tic marks are separated by half a unit on each
axis).}
\label{fig:k012}
\end{center}
\end{figure}

\begin{figure}[htb!]
\begin{center}
\includegraphics[width = 0.7\textwidth, angle=-90]{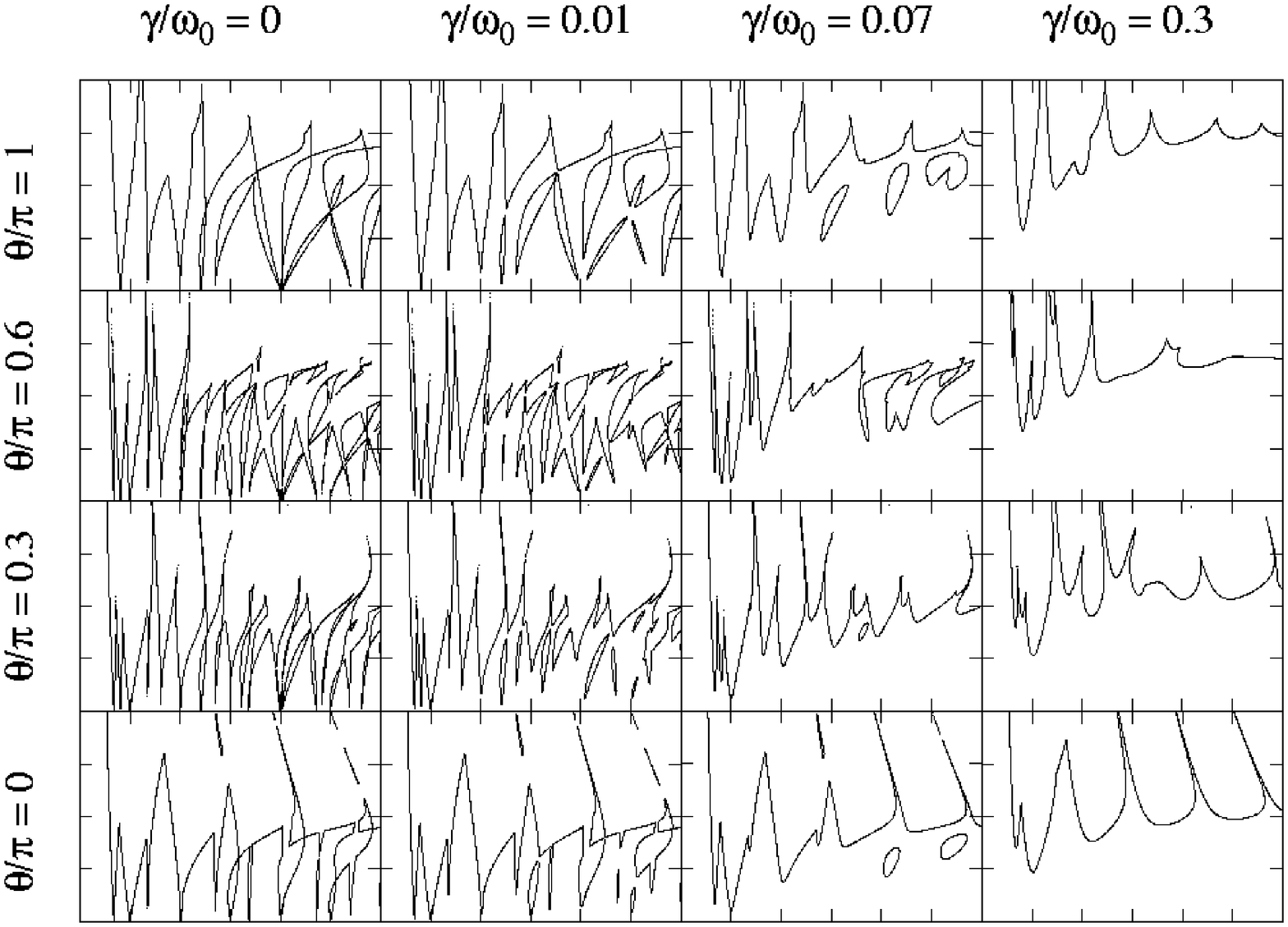}
\caption{$(r,A)$ plane as in Fig.~\ref{fig:theta0} for $k/\omega_{0}^2
= 0.6$ and several combinations of $\theta$ and $\gamma$ (origin is at
bottom left corner; tic marks are separated by half a unit on each
axis).}
\label{fig:k06}
\end{center}
\end{figure}

Perhaps the most striking feature of the $\theta$-dependence is its
sensitivity: the unstable regions in the $(r,A)$ plane can change very
abruptly (yet continuously) with the phase difference. As one might
expect, this sensitivity is primarily modulated by the coupling $k$,
as exemplified in Fig.~\ref{fig:gamma0} in the absence of damping.
The left and right columns of Fig.~\ref{fig:gamma0} show the
instability regions in the $(r,A)$ plane for $k/\omega_{0}^2 = 0.05$
and $k/\omega_{0}^2 = 0.5$, respectively.  The evolution of the
boundaries is shown as $\theta$ increases vertically along the columns
(in order to save space, this and subsequent figures consistently omit
labels, using the same scale as Fig.~\ref{fig:theta0} in the $(r,A)$
plane, with the origin at the bottom left corner and tic marks
separated by half a unit on both axes). While the most visible effects
of the phase difference seem to concentrate on the low $A$ region for
the smaller coupling (left column), the larger coupling (right column)
induces a richer behavior, with (in)stabilities arising also for
larger values of $A$ as $\theta$ changes. The phase difference can
therefore create regions of stability and instability which are absent
for in-phase modulations.

In order to address the effect of damping in this scenario, we have
chosen to fix the coupling $k$ and tile several $(r,A)$ instability
plots in such a way that $\theta$ varies in the vertical direction
while $\gamma$ varies in the horizontal direction. Fig.~\ref{fig:k012}
shows such a panel for a relatively small coupling $k/\omega_{0}^2 =
0.12$. One notices that the $\theta$ dependence is smoothed out as
damping increases, so that almost no structure is visible 
along the rightmost column ($\gamma/\omega_{0} =
0.3$). Fig.~\ref{fig:k06} shows the results for a larger coupling
$k/\omega_{0}^2=0.6$. Note that the same tendency is observed: the
instability boundaries become gradually less sensitive to $\theta$ as
$\gamma$ increases. However, the rightmost column of
Fig.~\ref{fig:k06} now shows much more structure than that of
Fig.~\ref{fig:k012}, indicating that for the same value of $\gamma$,
the $(r,A)$ plane with larger $k$ shows a stronger $\theta$
dependence.

\begin{figure}[htb!]
\begin{center}
\includegraphics[width = 0.8\textwidth, angle = -90]{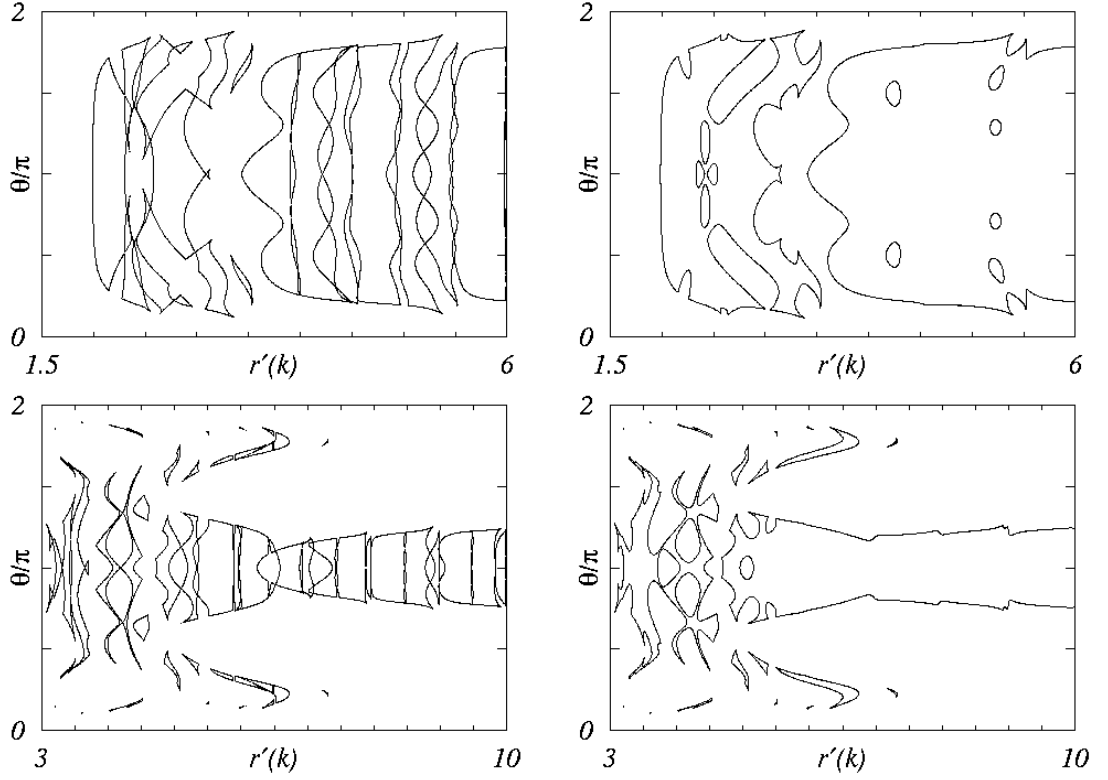}
\caption{Stability regions in the $(k,\theta)$ plane in the absence
(left column, $\gamma=0$) or presence (right column,
$\gamma/\omega_{0} = 0.05$) of damping. Top panels: $r= 1.9$ and
$A=0.9$. Bottom panels: $r = 2.25$ and $A= 1.3$.}
\label{fig:ktheta1}
\end{center}
\end{figure}

\begin{figure}[!hbt]
\begin{center}
\includegraphics[width = 0.8\textwidth, angle = -90]{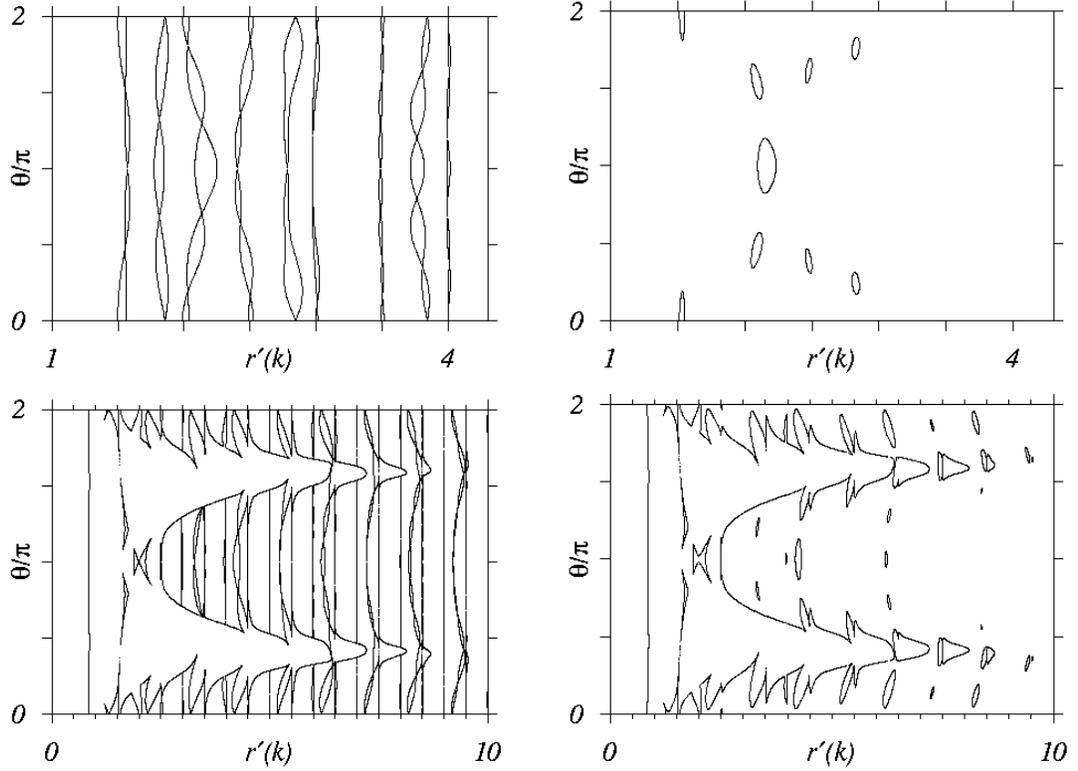}
\caption{Instability regions in the $(k,\theta)$ plane in the absence
(left column, $\gamma=0$) or presence (right column,
$\gamma/\omega_{0} = 0.05$) of damping. Top panels: $r= 1.25$ and
$A=0.6$. Bottom panels: $r = 0.8$ and $A= 1.8$.}
\label{fig:ktheta2}
\end{center}
\end{figure}

Figs.~\ref{fig:k012} and~\ref{fig:k06} suggest that $\theta = \pi$
gives rise to particularly stable behavior.  This is in agreement with
the vague intuitive notion that anti-phase modulations should be less
prone to resonance than in-phase ones. In order to verify the extent
to which this intuition is correct, a complementary view of these
phenomena can be obtained by projecting the instability regions in the
$(k,\theta)$ plane instead. This allows us to start with uncoupled
oscillators ($k = 0$) and observe how a given (in)stability evolves as
$k$ and $\theta$ change. Indeed, it turns out to be possible to
understand much of the behavior of the coupled system in terms of the
behavior of the uncoupled system.  To produce the representative
results shown in Figs.~\ref{fig:ktheta1} and~\ref{fig:ktheta2}, we fix
several $(r,A)$ points in the single oscillator stability boundary
diagram shown as black circles in the first panel of
Fig.~\ref{fig:theta0} and study the way in which variations in $k$ and
$\theta$ affect these particular states.  Two of the points in
Fig.~\ref{fig:theta0}, $(r,A) = (0.8,1.8)$ and $(1.25,0.6)$ are stable
states for the single oscillator (the first black dot touches the
stability boundary in the figure only because it has been drawn large
enough to render it visible; the point is well within the stable
region).  The other two points, $(r,A)=(1.9.0.9)$ and $(2.25,1.3)$
lead to unstable behavior of the single oscillator.

The first thing to be noted in Fig.~\ref{fig:ktheta1} is that the
horizontal axis has been rescaled in order to reveal the relevance of
the variable $r^{\prime}$ (see Eq.~(\ref{eq:rprime})). The top panels
focus on $r= 1.9$ and $A=0.9$, which is in the parametrically resonant
regime for the uncoupled system.  The small-$r'$ portion of the figure thus
represents an unstable regime. One notes that for $r^{\prime}(k)$
slightly above 2 the system becomes stable within a $\theta$ interval
centered around $\pi$. Increasing $k$ a little further, this stability
region then evolves in a very complex pattern, which includes
reentrant ``holes'' of instability. After a distinguishable gap of
instability, a somewhat simpler band of stability arises around
$r^{\prime}\sim 3.5$ starting at $\theta=\pi$.
This band continues to larger values of $k$, with
its outermost boundaries presenting a relatively simpler envelope than the
low-$k$ pattern. The interesting point to be emphasized is that
the band is perforated by gaps of instability most of which
are centered precisely at integer and half-integer values of
$r^{\prime}$. This result is perhaps
anticipated by the fact that the frequency
$\omega_0^\prime$ appears as the effective average diagonal frequency in the
mean field equation of motion in~\cite{Bena99b}).  The gaps
are eventually closed by increasing the damping (top right panel),
which also broadens the stability band and simplifies its dependence
on the phase difference.  The bottom panels in Fig.~\ref{fig:ktheta1}
are for $r = 2.25$ and $A= 1.3$, which again is in the resonant
regime of the single oscillator. One notices the same pattern: for
lower values of $k$, a complex shape emerges in the $\theta$
dependence of the stable regions. For sufficiently large values of
$k$, a band of stability arises which has instability gaps basically
centered at integer and half-integer values of $r^{\prime}$. Damping
(bottom right panel) causes gaps to disappear, creating
a uniform region of stability centered around $\theta=\pi$.

Fig.~\ref{fig:ktheta2} presents what may be regarded as the opposite
situation, namely, when the original uncoupled system is stable. The
top panels show the results for $r = 1.25 $ and $A = 0.6$. Notice that
the behavior is much simpler in this case, with the original stability
being disturbed mostly around integer and half-integer values of
$r^{\prime}$ in the absence of damping (top left), with a relatively
weak $\theta$ dependence. The effect of damping (top right) is to
suppress most of these instability regions, yielding a predominantly
stable $(k,\theta)$ plane. The bottom panels show results for an
interesting intermediate situation: even though the uncoupled system
is stable for $r = 0.8$ and $A = 1.8$, this point lies in a narrow
corridor between two instability regions in the $(r,A)$ plane (see
Fig.~\ref{fig:theta0}). One would therefore expect instabilities to
arise more easily, and the immediate question is how the resulting
diagram might reconcile the structures observed in
Fig.~\ref{fig:ktheta1} and the top panels of
Fig.~\ref{fig:ktheta2}. The answer lies in a very rich structure in
the $(k,\theta)$ plane (bottom left panel): initially, very small
coupling induces an instability for all $\theta$.
The now predominantly unstable system evolves in a manner
similar to those of Fig.~\ref{fig:ktheta1}, a stability region being
created around $\theta = \pi$ as $k$ increases, with string-like gaps
of instability around the usual values of $r^{\prime}$. The difference
is that there is now also a second band of stability centered around
$\theta = 0$, presenting gaps at the same $r'$ positions. For $k$
sufficiently large, those two stability bands merge and one is left with a
structure similar to that of the top panels: a predominantly stable
region permeated by strings of instability. The bottom right panel of
Fig.~\ref{fig:ktheta2} shows the effect of damping, which
significantly reduces the instability gaps thus greatly simplifying the
picture.

Therefore the notion that modulations operating with a phase
difference $\theta = \pi$ are less prone to parametric resonance is
thus seen to be essentially correct. Framing this statement more
carefully, our results show that for given $r$ and $A$, sufficiently
large values of $k/\omega_{0}^2$ and $\gamma/\omega_{0}$ are able to
induce, in the $(k,\theta)$ plane, a band of stability centered around
$\theta=\pi$ even if the uncoupled oscillators are individually
unstable.  The width of this band can eventually comprise the whole
$2\pi$ interval if the uncoupled system is originally stable.

\section{Collective parametric instability}
\label{sec:collective}

Bena and Van den Broeck~\cite{Bena99b} studied the stability boundaries
of $N$ parametrically modulated oscillators $\{x_{i}\}$ each
coupled to all the others by the same coupling constant $2k/N$ (in our
notation). The phases $\{\theta_{i}\}$ are initially
chosen at random from a uniform distribution in the interval
$[0,2\pi]$, remaining quenched thereafter.
With square-block modulation the system is exactly solvable
in the limit $N\to \infty$, where the mean-field solution becomes
exact. The mean field equation is 
\begin{equation}
\label{eq:meanfield}
\ddot{x} =  -\omega_{0}^2[1+\phi_\theta(t)]x -\gamma \dot{x}
- 2k(x-\left< x\right>)
\end{equation}
where $x$ is the displacement of any oscillator in the chain,
$\phi_\theta(t)$ is the periodic
modulation with phase $\theta$, and 
$\left<x\right> \equiv N^{-1} \sum_{i=1}^{N}x_{i}$ is the mean displacement
to be determined self-consistently.  Bena and Van den Broeck note that
the exact solution of this equation is
\begin{equation}
\label{eq:meanfieldsol}
\left( \begin{array}{c} x(t) \\
\dot{x}(t) \end{array} \right) = \mathbf{G}_\theta (t) \cdot \left(
\begin{array}{c} x(0) \\ \dot{x}(0)\end{array}\right) + 2k  \mathbf{G}_\theta
(t) \cdot \int_0^t d\tau  \mathbf{G}_\theta (\tau)^{-1} \cdot 
\left( \begin{array}{c} 0 \\ \left< x(\tau)\right>
\end{array} \right) \; .
\end{equation}
The propagator $\mathbf{G}(t)$ is known explicitly. Indeed, at
$t=T$ it is $\mathbf{G}(T) = e^{-\gamma T} \hat{F}(T)$ where
$\hat{F}$ is precisely the single-oscillator Floquet operator given in
Eqs.~(\ref{eq:floquet1}) and (\ref{eq:floquet2}), but now with the
frequencies shifted by the coupling constant
\begin{equation}
\label{eq:shifted}
\omega_\pm = \sqrt{\omega_0^2(1\pm A) +2k -\gamma^2/4}.
\end{equation}
Note that this is {\em exactly} the same as the propagator associated with
the $\rho$ variable of Eq.~(\ref{eq:rho}) in the two-oscillator in-phase
modulation problem, that is, the propagator associated with a single
oscillator of frequency $\omega_0^\prime$.  

\begin{figure}[hbt!]
\begin{center}
\includegraphics{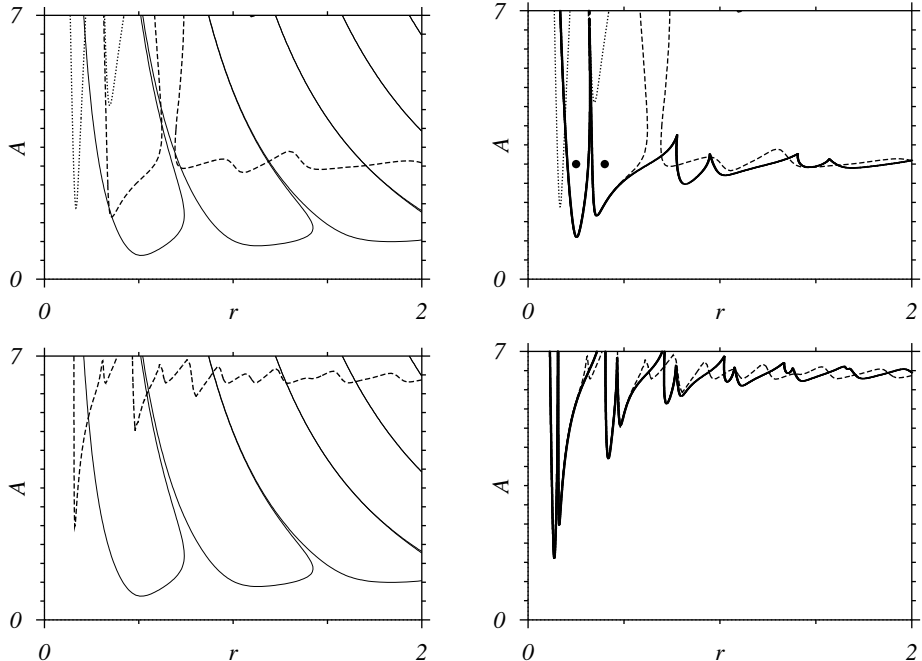}
\caption{Instability regions in the $(r,A)$ plane for
$\gamma/\omega_{0} = 0.4$. According to the mean field 
model~\cite{Bena99b}, usual parametric resonance occurs inside
the dotted boundaries, while the boundaries of collective instability
are depicted by dashed lines. $k/\omega_{0}^{2} = 4$ for the top
panels, and $k/\omega_{0}^{2} = 20$ for the bottom panels. The thin
solid lines correspond to $k = 0$ (left column); the thick solid lines
(right column) correspond to $\theta = \pi$ in the two-oscillator
model. The dark points in the upper right panel are parameter pairs
considered in more detail subsequently.}
\label{fig:collective}
\end{center}
\end{figure}

Bena and Van den Broeck identify two sorts of instabilities.  One, which
they call the ``usual parametric resonance," arises from the divergence
associated with eigenvalues of $\mathbf{G}$ of magnitude greater than
unity, that is, with the unbounded growth of the first term in
Eq.~(\ref{eq:meanfieldsol}) which in turn signals the unbounded growth of
the amplitude of any typical oscillator in the chain. 
The stability boundaries associated with this type of instability
are given precisely by Eq.~(\ref{eq:boundary1}) and are shown for
the parameter choices indicated in the caption
as the dotted curves in the top row panels of Fig.~\ref{fig:collective}.  
In our in-phase two-oscillator parlance
these are exactly the boundaries of the ``$r'$-instability regions" 
defined in terms of the shifted frequency $\omega_0^\prime$ (cf.
Eq.~(\ref{eq:omegaprime})). 
The ``usual" regions shrink in width and move toward lower $r$ and larger
$A$ with increasing coupling $k$, a behavior already exhibited in the
context of the in-phase two-oscillator results of Fig.~\ref{fig:theta0}.
Indeed, this instability is beyond the scale of the figures in the
large-coupling bottom row panels.
The other type of instability, which they call a ``collective instability,"
is associated with the divergence of the mean $\left< x \right>$ and
hence of the second term in Eq.~(\ref{eq:meanfieldsol}).  The collective
instability boundaries are shown as dashed curves in all panels of
Fig.~\ref{fig:collective}. Note that the two types of instabilities may
occur simultaneously, as seen in the instability region overlap
evident in the top row panels of the figure.  We return to this point below.
Note also that with increasing coupling the system becomes increasingly
stable, as one might expect, and that the unstable behavior becomes
primarily collective.

We wish to explore whether our two-oscillator model is able to capture at
least some of the behavior of the infinite system.  In particular, we would
like to investigate whether features of the collective instabilities
of the mean field model are apparent in a system of only two oscillators
with a fixed relative modulation phase. To make the comparison we also show
in Fig.~\ref{fig:collective} the stability boundary for a single uncoupled
oscillator (thin solid lines in left column) and for coupled oscillators 
with relative modulation phase $\theta=\pi$.

\begin{figure}[hbt!]
\begin{center}
\includegraphics{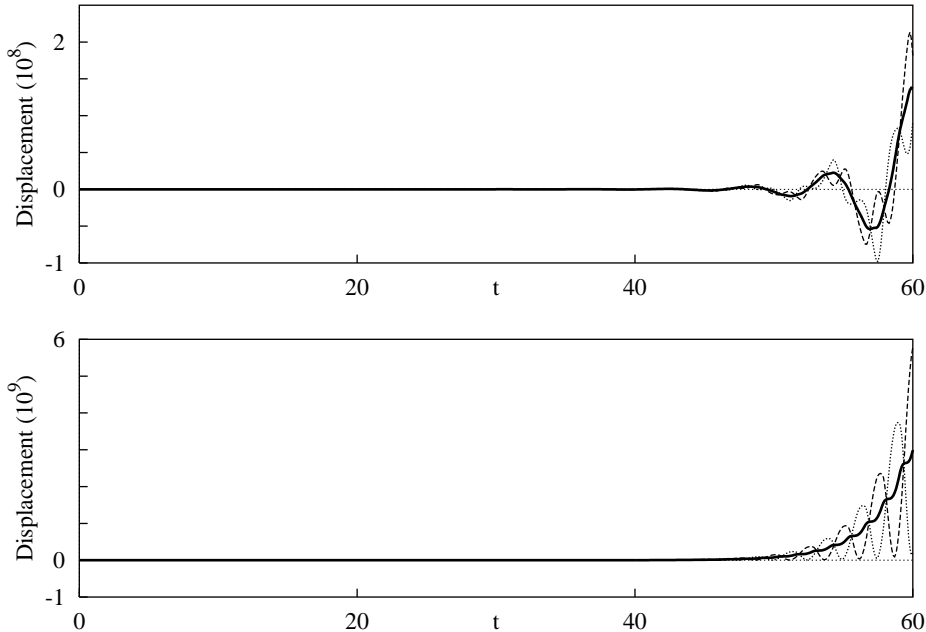}
\caption{Trajectories for the two-oscillator $\theta=\pi$ system for
the parameters corresponding to the dark points in the upper right panel
of Fig.~\ref{fig:collective}: 
$\gamma/\omega_0=0.4$, $A=3$, and $k/\omega_0^2=4$.  $r=0.25$ in the upper
panel and $0.4$ in the lower panel.  The dashed curves are the displacement
$x_1$ of one oscillator vs time, the dotted curves are the displacement
$x_2$ of the other oscillator, and the thick solid curves represent
the mean displacement $x=(x_1 + x_2)/2$.  
}
\label{fig:trajectory}
\end{center}
\end{figure}

We make the following assertions: the two-oscillator system
with {\em any} value of $\theta$ captures
features of the overlap region of ``usual" and ``collective"
instabilities.  The {\em purely} ``usual" regions are captured most
accurately by the $\theta=0$ system, and the {\em purely}
``collective" regime is
best captured by the two-oscillator system with $\theta=\pi$.  It is
therefore this latter system that most fully captures 
(with unexpected detail) the principal features of collective behavior of
the mean field model, and does so with increasing accuracy as the coupling
between oscillators increases.  We support these assertions,
particularly the last one which is the one of most interest to us,
with the results shown in Figs.~\ref{fig:collective}
and \ref{fig:trajectory}.

Clearly, the $\theta=0$ system captures the full ``usual" instability regime
{\em exactly} since, as already stated, the ``usual" instability is exactly
the same as the ``$r'$-instability." This identity is not restricted to the
square wave modulation but holds for any periodic modulation.  
However, the $\theta=0$ system does not capture the ``collective"
instability since the ``$r$-instability" condition is that
associated with a single parametric oscillator with the unshifted frequency
$\omega_0$.  Thus, for example, in the
first panel of Fig.~\ref{fig:collective} the $\theta=0$ instability
boundaries can be constructed from the combination of the dotted regimes
and the thin solid lines (compare with the right lower panel of
Fig.~\ref{fig:theta0}), whereas those of the mean field system include
the same dotted regimes but now the very different dashed regions.
The left lower panel shows an even greater difference between the $\theta=0$
two-oscillator model (whose ``$r$-instability" boundaries are independent
of $k$) and the mean field model (where the boundaries of the collective
instability are sensitively dependent on $k$).  Two coupled parametric
oscillators with relative modulation phase $\theta=0$ therefore
do not capture the collective features of the mean field model.

Our assertion that the two-oscillator system with any $\theta$ contains
elements of the ``usual" instabilities in the mean field model is simply a
restatement of our earlier observation that ``$r'$-instabilities"
continue to appear even when one moves away from $\theta=0$ and all the way
to $\theta=\pi$.

Consider now the two coupled oscillators with $\theta=\pi$. The stability
boundaries  are shown by the thick solid lines
in the panels in the right column of
Fig.~\ref{fig:collective}.  In the upper panel
we observe that the first tongue approximates the region of ``usual" mean
field instability (in the non-overlapping regime)
and that the remainder captures the collective
instability boundary features surprisingly well (although it does miss
the gap). We particularly point to the excellent fit of the leftmost
boundary of this region.  The agreement between the two models is
even more dramatic in the lower panel, which corresponds to stronger
coupling $k$.  Again, the details are surprisingly well matched and the
leftmost boundary of the region is captured essentially exactly.

To further support our analysis, and to gain a clearer 
understanding of the difference between ``usual" and ``collective" 
instabilities (which are both evidently already present in our 
two-oscillator system although the notion of a collective effect 
is not obvious in such a small system), we consider the 
motions that might characterize the instabilities.  In the mean 
field system we conjecture that in the non-overlapping ``usual" 
instability regime the mean is zero, $\left< x \right>=0$, but 
each oscillator oscillates about zero with ever increasing 
amplitude.  The motion in the non-overlapping ``collective" 
instability regime may involve an ever increasing mean with each 
oscillator oscillating about this moving mean with finite 
amplitude. This description is in accord with that of Bena and Van 
den Broeck~\cite{Bena99b}. The overlap regions may involve both an 
increasing mean and oscillations of ever increasing amplitude 
about the moving mean.  We have not ascertained these conjectures 
in the mean field system, but present results for the 
two-oscillator system that support this description. 

Figure~\ref{fig:trajectory} shows trajectories for the two-oscillator
$\theta=\pi$ system at the two points marked on the upper right panel
of Fig.~\ref{fig:collective}.  The trajectories shown are those of
each of the two oscillators as well as the mean trajectory.  The upper
panel is for parameter values in the unstable region that is {\em not}
in the ``collective" regime.  It is tempting to associate this with
the non-overlapping ``usual" instability of the mean field model, an
association that requires some caution.  The figure indicates that not
only does each oscillator and also the mean oscillate about zero, but
all the trajectories, {\em including} the mean, appear to diverge.
This behavior is that envisioned in our earlier discussion of the
$\theta=0$ two-oscillator system in the regime where
``$r$-instabilities" and ``$r'$-instabilities" overlap, and is an
indication that features of both kinds of instabilities persist even
at $\theta=\pi$.  We conjecture that ``$r$-instabilities" are finite
size divergences not present in the mean field model (and hence we
would not expect to see this particular type of trajectory there). In
the mean field limit only the ``$r'$-instability" contributions in
this regime persist, becoming the non-overlapping ``usual" instability
portion of the phase boundary portrait.

The lower panel of Fig.~\ref{fig:trajectory} is for parameter values
in the non-overlapping ``collective" regime.  The mean indeed increases,
and each oscillator oscillates about this increasing mean.  It is
interesting that the main feature of the ``collective" instability, namely,
an increasing mean displacement with individual oscillators oscillating
about this mean, is already clearly captured by the two-oscillator model.

\section{Conclusions}
\label{sec:conclusions}

We have presented results for a model of two coupled
parametric oscillators that can be solved for any value of the phase
difference $\theta$ between the modulations. The other parameters
of the problem are the oscillator frequency $\omega_0$, the modulation
amplitude $A$ and period $T=2\pi/\omega_p$, the coupling $k$ and 
the damping $\gamma$. 

We found that in-phase ($\theta=0$) coupled oscillator motion can be 
separated into two independent contributions.  One
involves the two oscillators moving
together about zero (``$r$-instabilities").  The instability 
boundaries for this motion are identical to those of a single 
parametric oscillator of frequency $\omega_0$ and are
independent of coupling $k$ since the spring connecting the 
oscillators is never disturbed. 
The other involves the two oscillators moving about zero but exactly
in anti-phase with one another (``$r'$-instabilities"). The 
stability boundaries are 
sensitive to $k$ for these motions: the system becomes more stable 
with increasing coupling.  We noted that these latter 
instabilities are exactly those identified as ``usual"
instabilities in the mean field model~\cite{Bena99b} and that they
contribute to the instability boundaries in our two-oscillator 
model for any $\theta$, not just for $\theta=0$.  We also showed 
that damping shrinks the instability regimes and 
smoothes the stability boundaries.

We showed that a change in $\theta$ can substantially modify the
regions of parametric instability as projected in the usual
$(r=\omega_{0}/\omega_p, A)$ plane, and that these changes are
strongly affected by the coupling between the oscillators.  In
general, increasing $\theta$ up to $\pi$ provides greater stability
but also leads to more intricate stability boundaries.  An increase in
$k$ and/or $\gamma$ also leads to increased stability.  Alternatively,
it is possible to understand the coupled system in terms of the
uncoupled system by projecting the instability regions onto the
$(k,\theta)$ plane.  This rendition is particularly useful to show
that $\pi$-centered bands of stability arise for sufficiently large
$k$ and $\gamma$.  We have thus identified all the trends of behavior
in the two-oscillator model as each of the parameters is varied.

Our most interesting insights arise from a 
comparison of the two-oscillator results with the mean field 
model~\cite{Bena99b} consisting of $N\rightarrow\infty$ 
parametrically modulated oscillators all coupled to one another.
In this latter system two types of instabilities have been
identified: ``usual" and ``collective."  The ``usual" 
instabilities are exactly our ``$r'$-instabilities," that is,
the stability boundaries are identical.  The interesting
result is that the ``collective" instabilities already appear in 
the two-oscillator model with $\theta=\pi$.  This statement
is based on the great similarity of the stability boundaries
of the mean field and two-oscillator systems, especially with
increasing coupling, and the specific features of the oscillator 
trajectories that typify the motions in each of these unstable 
regimes.  It is perhaps surprising that a two-oscillator
model can capture so much of the mean field collective behavior,
and suggests that collective resonance in the latter 
may be dominated by phases quenched around $\pi$.

\vspace*{0.5in}
%\vskip 12pt
\appendix{\bf APPENDIX A - Parametric Resonance for a Single Oscillator}
\label{a}
\\
\setcounter{equation}{0}
\newcounter{appendix}
\setcounter{appendix}{1}
\renewcommand{\theequation}{\Alph{appendix}.\arabic{equation}}
A brief review of parametric resonance for a single oscillator provides
results called upon in the body of the paper for the coupled
system~\cite{Arnold,Bena}.  The equation of motion is
\begin{equation}
\label{eq:oneosc}
\ddot{x}+\omega^2(t)x +\gamma \dot{x} = 0\; .
\end{equation}
The frequency $\omega(t)$ is a periodic function with period $T \equiv
2\pi/\omega_{p}$. The simplest model of such a parametric modulation is
perhaps the square wave
\begin{equation}
\omega^{2}(t) = \omega_0^{2}\left[1+A\mbox{sgn}
(\sin(\omega_{p}t))\right]\; .
\end{equation}
In the undamped case ($\gamma=0$) the oscillator frequency thus switches
periodically between the two values $\omega_{\pm}=\omega_0\sqrt{1\pm A}$.
If $A\leq 1$, the system jumps between two oscillatory regimes, each
of which is stable. If $A>1$, then $\omega_{-}$ becomes
purely imaginary, corresponding to a saddle node behaviour in the
phase space $(x,\dot{x})$. In this case, the system jumps between an
unstable saddle and a stable oscillatory regime.  As we shall see below,
both instability and stability can occur for either $A\leq 1$ or $A>1$.
With damping, the relevant frequencies are shifted (see below).
The stability of the system depends on the interplay between the
amplitude $A$ of the modulation and its frequency $\omega_p$ as well as
the system parameters $\omega_0$ and $\gamma$.

The linearity of the equations leads naturally to the application of
Floquet theory for the solution of the problem~\cite{Nayfeh}.
A change of variables to
\begin{equation}
\label{eq:newvar}
y(t)= e^{\gamma t/2} x(t)
\end{equation}
yields the undamped equation
\begin{equation}
\label{eq:oneoscmod} \ddot{y}+\tilde{\omega}^2(t)y = 0\; 
\end{equation}
where
\begin{equation}
\label{eq:modfreq}
\tilde{\omega}^2(t) = \omega^2(t)-\frac{\gamma^2}{4}.
\end{equation}
For the square wave modulation we then have
\begin{equation}
\tilde{\omega} =
\left\{
\begin{array}{l}
\tilde{\omega}_{+} \equiv \sqrt{\omega_0^2(1+A)-\gamma^2/4} \\ [12pt]
\tilde{\omega}_{-} \equiv \sqrt{\omega_0^2(1-A)-\gamma^2/4}
\end{array}
\right.
\end{equation}
 
Due to the periodicity of $\tilde{\omega}(t)$ the two 
independent solutions of Eq.~(\ref{eq:oneoscmod}) must also have 
period $T$. One can write for the original variable 
\begin{equation}
\label{eq:genfloquet}
\mathbf{X}(t+T) =  e^{-\gamma T} \hat{F}(T) \mathbf{X}(t)\; ,
\end{equation}
where
\begin{equation}
\mathbf{X}\equiv \left(\begin{array}{c}x\\
\dot{x}\end{array}\right)
\end{equation}
and $\hat{F}$ is the so-called Floquet operator
that (aside from damping) propagates the system in phase space
for one period:
\begin{equation}
\label{eq:floquet1}
\hat{F}(T)  = \hat{f}_{-}\left(T/2\right)\hat{f}_{+}\left(T/2\right)\; .
\end{equation}
where 
\begin{equation}
\label{eq:floquet2}
\hat{f}_\pm (t)=
\left(\begin{array}{cc}
\cos \tilde{\omega}_\pm t + \frac{\gamma}{2\tilde{\omega}_\pm}
\sin \tilde{\omega}_\pm t &
\frac{1}{\tilde{\omega}_\pm} \sin \tilde{\omega}_\pm t \\ [12pt]
\-\frac{\omega_\pm^2}{\tilde{\omega}_\pm} \sin \tilde{\omega}_\pm t&
\cos\tilde{\omega}_\pm t - \frac{\gamma}{2\tilde{\omega}_\pm}
\sin \tilde{\omega}_\pm t 
\end{array}\right) 
\end{equation}
Equation~(\ref{eq:genfloquet}) implies that
\begin{equation}
\mathbf{X}(t+nT) = e^{-n\gamma T} \hat{F}^{n}(T)\mathbf{X}(t)\; ,
\end{equation}
so the long term behavior of the system is governed by the eigenvalues
of $e^{-\gamma T}\hat{F}$. If any eigenvalue is larger than 1
in absolute value, 
there is parametric resonance and the system is unstable.

One can show that the two eigenvalues of $e^{-\gamma T}\hat{F}(T)$
obey the relation
$\lambda_{1}\lambda_{2} = 1$, reflecting the incompressible piecewise
Hamiltonian flow. Therefore the eigenvalues are complex conjugates in
the stable regions and real in the unstable
regions.  At the boundaries both
eigenvalues equal either $+1$ or $-1$, in which case the system oscillates
with period $T$ or $2T$, respectively. Therefore the trace of the
Floquet operator is $\pm 2$ at the stability boundaries, leading to the
following boundary equation:
\begin{equation}
\label{eq:boundary1}
\cos\left(\frac{\tilde{\omega}_{-}T}{2}\right)
\cos\left(\frac{\tilde{\omega}_{+}T}{2}\right) -
\left(\frac{\tilde{\omega}_{+}^2+\tilde{\omega}_{-}^2}
{2\tilde{\omega}_{+}\tilde{\omega}_{-}}\right)
\sin\left(\frac{\tilde{\omega}_{-}T}{2}\right)
\sin\left(\frac{\tilde{\omega}_{+}T}{2}\right) =
\pm \cosh\left(\frac{\gamma T}{2}\right)\; .
\end{equation}

In the absence of damping the solutions depend only on $A$ and the ratio
$r\equiv \omega_{0}/\omega_{p}$, and these particular results are shown by
the dashed curves in the first panel of Fig.~\ref{fig:theta0}.
The regions of instability appear as ``tongues'' starting from integer and
half-integer values of $r$ at small modulation amplitudes.  Note that
the stability boundaries surrounding the tongues that emerge from integer
values of $r$ correspond to both Floquet eigenvalues being equal to $1$,
while those surrounding tongues that emerge from half-integer values of
$r$ are associated with the Floquet eigenvalues equal to $-1$.
Note that no abrupt transition (nor a visible transition of any kind) is
seen at the line $A = 1$ which marks the transition between an
oscillating-oscillating behavior and a saddle-oscillating one. As one
would expect, on the other hand, the system is nearly always unstable
for sufficiently small $\omega_{p}$ and sufficiently large $A$ (upper
right corner of the figure).

\paragraph{Acknowledgements:}
We would like to thank C. Van den Broeck and R. Kawai for stimulating
discussions. This work was supported in part by the 
National Science Foundation under grant No. PHY-9970699.

%\bibliography{/home/bassi/mauro/latex/copelli}

\end{document}